\documentclass[twocolumn,amsmath,amssymb,aps,pra,floatfix,nofootinbib,superscriptaddress]{revtex4}
\usepackage{graphicx,graphics,times,color}% Include figure files
\usepackage{bm}% bold math
\usepackage{longtable}

\def\be{\begin{equation}}
\def\ee{\end{equation}}
\def\ba{\begin{eqnarray}}
\def\ea{\end{eqnarray}}
\def\la{\langle}
\def\ra{\rangle}

\begin{document}

\title{Spin State Transfer in Laterally Coupled Quantum Dot Chains with Disorders}
\author{Song Yang}
\affiliation{Department of Physics and Astronomy, University College
London, Gower St., London WC1E 6BT, United Kingdom}
\affiliation{Key Laboratory of Quantum Information, University of
Science and Technology of China, Hefei, 230026, People's Republic of
China}
\author{Abolfazl Bayat}
\affiliation{Department of Physics and Astronomy, University College
London, Gower St., London WC1E 6BT, United Kingdom}
\author{Sougato Bose}
\affiliation{Department of Physics and Astronomy, University College
London, Gower St., London WC1E 6BT, United Kingdom}

\date{\today}

\begin{abstract}
Quantum dot arrays are a promising media for transferring quantum information between two distant points without resorting to mobile qubits.
Here we study two most common disorders namely, hyperfine interaction and exchange coupling fluctuations,
in quantum dot arrays and their effects on quantum communication through these chains.
Our results show that the hyperfine interaction is more destructive than the exchange coupling fluctuations.
The average optimal time for communication is not affected by any disorder in the system and our simulations show
that anti-ferromagnetic chains are much more resistive than the ferromagnetic ones against both kind of disorders.
Even when time modulation of a coupling and optimal control is employed to improve the transmission, the anti-ferromagnetic chain performs much better.
We have assumed the quasi-static approximation for hyperfine interaction and time dependent fluctuations
in the exchange couplings. Particularly, for studying exchange coupling fluctuations we have considered the static disorder,
white noise and $1/f$ noise.
\end{abstract}

%\pacs{03.65.Wj}
\maketitle

\section{introduction}
The transmission of quantum information between two well separated parties via
quantum channel is prerequisite for quantum communication and scalable quantum computation.
Spin chains are of great interest in quantum information science since they are natural candidates for quantum channels
\cite{bose03} in atomic scales. In the use of spin chains for quantum communication a sender can send a quantum state or share entanglement
with another separated set of spins at a
distant point of the spin chain just through the natural evolution of the system. Besides controlling the sender and
the receiver spins, no extra controls are needed for communication so that system can be shielded from the environment
to minimize decoherence. Based on the physical implementation of spin chains several imperfections can affect
the communication process. Thermal fluctuations \cite{bayat-thermal} and decoherence \cite{burgarth-dec, bayat-AFM, bayat-xxz} have
been studied as external effects. Another important source of imperfection is disorder which is inevitable due to imperfect
fabrication processes. In any physical implementation, always there exist some parameters which cannot be tuned perfectly.
For instance in a spin chain one cannot guarantee to have precise couplings without disorder and also each spin can have different energy
splitting due to a fluctuating electric or magnetic field.
In \cite{Petrosyan2} the coherent dynamics of one and two electron transport in a linear array of tunnel-coupled
quantum dots in the presence of imperfect fabrications have been studied.  
Moreover, the influences of the static disorders on XX spin chain model have been analyzed recently \cite{david10} and was shown that
locally controlling the couplings is more susceptible to disorders than permanently coupled chains.
On-site energy fluctuations in spin chains have been considered in \cite{Eisert} and it was found that these fluctuations suppress the
transmission in a different way compared to the static disorders.
Due to the random nature of disorder, they also may cause localization in long chains which
restricts the communication length. This localization and communication beyond that length has been investigated in \cite{Linden}.

Chains of perpetually coupled spins or other qubits in solid state systems,
may be used to connect solid state quantum registers without resorting to optics. Thus, proposals with chains of charge qubits \cite{Romito-fazio-bruder},
flux qubits \cite{lyakhov-bruder} and quantum dot based excitonic qubits \cite{ami07,spiller07} have been put forward.
However, in this context spins in quantum dot arrays, look particulary promising,
since electron spins in quantum dots have relatively long relaxation time \cite{Elzerman04,taylor07,hanson03,krout04},
allow for coherent manipulations \cite{ronald08,petta05,nowack07,koppen05}.
They will be ideal as connectors between quantum registers built with spin qubits in quantum dots \cite{daniel98,burk99,hanson-burkard}.
The other advantage of using quantum dot arrays for realization of quantum channel is the easy and flexible
manipulation of the exchange couplings between neighboring dots. Theoretical \cite{burk99}
works has shown that the quantum dot chain might fairly easily transit from ferromagnetic (FM) to
anti-ferromagnetic (AFM) phase by modulating the barrier of neighboring dot or external magnetic field and
typically the interaction is found to be anti-ferromagnetic \cite{petta05}.

It has been shown that the perfect state transfer can be achieved in a chain of spins interacting permanently through
engineered couplings \cite{christandl04} or controlling a single local actuator which modulates
one energy-level transition \cite{schirmer08} in an XX Hamiltonian .
However, in the chain of quantum dots the natural interaction between neighboring spins is Heisenberg Hamiltonian \cite{daniel98} and
there is no way to convert it to a XX Hamiltonian for achieving perfect state transfer. On the other hand it was shown that 
in the Heisenberg Hamiltonian without locally modulating the magnetic field one cannot achieve perfect state transferring \cite{Marcin08}.

For electron spins in a mesoscopic  open quantum system, the most significant interactions are the
spin-orbit and the hyperfine interactions \cite{hanson07}. The first process can be efficiently
suppressed via reducing the temperature and also its time scale is so long such that for a fast
coherent scheme, such as state transferring, it does not have a significant effect. So, as the
first important effect in quantum dot spin chain communication we focus on the hyperfine interaction which
practically can not be suppressed due to the permanent interaction with the spins of nuclei in the host material.

Moreover, having a strong spin exchange coupling, for a fast evolution, by means of external gates will introduce
background charge fluctuations in the system. This charge fluctuation will induce variations of
spin exchange coupling, which also lead to qubit dephasing. Unlike the hyperfine interaction
the quasi static approximation is not valid for exchange coupling fluctuations and they suffer from a time dependent disorder which behaves like $1/f$ noise \cite{chris04}.

In this paper, we study the effect of hyperfine interaction and exchange coupling fluctuation over the quality of quantum communication
through the quantum dot spin chains. We consider linear lateral quantum dot arrays
in both FM  and AFM regimes and compare the destructive effects of these two source of imperfections on the quality of communication.
As hyperfine interactions lead to non-conservation of total magnetization of the chain
here we require a general formula for the fidelity of quantum state transfer in an arbitrary quantum channel. Accordingly,
we present and use such a formula, which to our knowledge, has not been used in the spin chain literature.

The structure of the paper is as following. We first introduce the theoretical model to
realize the state transfer based on quantum dot arrays in Sec. \ref{method}. Then the effects of
hyperfine interaction and
exchange interaction fluctuation are
investigated in Sec. \ref{hyperfine} and Sec. \ref{fluctuation}. Moreover we investigate the quantum state transfer in practical situation including hyperfine interaction, exchange interaction fluctuation as well as thermal fluctuations in Sec. \ref{practical}.
A possible improving strategy via quantum control theory is discussed in Sec. \ref{optimal}. Finally, our conclusion follows in Sec. \ref{conclusion}.

\section{Quantum State transferring in an Ideal Chain without Disorder}
\label{method}
We consider a linear array of lateral GaAs quantum dots, electrostatically defined in a two-dimensional electron gas via metallic gates on the
top of a semiconductor heterostructures (GaAs/AlGaAs) \cite{hanson07,petta05}.
Here each dot is doped with a single excess electron, and qubit is encoded on
the electron spin. When tunneling barrier is ``high", the interactions between neighboring
dots are forbidden; and if tunneling barrier is ``low",
the spins will experience an exchange interaction which can be described by the Heisenberg
model \cite{daniel98}. An external magnetic field $h_{z}$ can be applied in the $z$ direction
to break the degeneracy between two spin levels,
i.e. $|0\ra=|\downarrow\ra$ and $|1\ra=|\uparrow\ra$ with a Zeeman
splitting $\Delta_{z} = g\mu_{B}h_{z}$.

\begin{figure} \centering
    \includegraphics[width=8cm,height=7cm,angle=0]{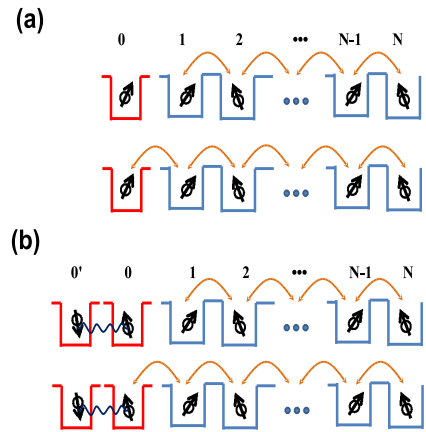}
    \caption{(Color online) (a) Scheme for transferring an arbitrary pure state through the quantum dot chain. Spin 0 is the sender qubit
and initially is decoupled from the channel qubits (spins $1, 2, ..., N$) which are prepared in their ground state. The sender places the quantum
 information on the 0th qubit, and switches on the interaction between the 0th and the 1st qubit of the channel in order
 to send the information to the $N$th qubit. (b) Scheme for entanglement distribution.
 At the beginning, the channel is initialized to its ground state, and a singlet state is prepared between
 spin 0 and spin $0^{'}$. The sharing entangled information propagates from the 0th spin to the $N$th one by
 switching on the coupling between the 0th spin and the 1st spin while spin $0^{'}$ remains decoupled from the rest of the system during the evolution.}
     \label{fig0}
\end{figure}

In Fig. \ref{fig0} (a) we have shown the schematic of the system. Spin $0$ is initially decoupled from the others while the rest of the system are interacting through the
following Hamiltonian
\begin{equation}
H_{ch}=\sum\limits_{k=1}^{N-1}J_{k}\hat{\textbf{S}}_{k}\cdot\hat{\textbf{S}}_{k+1},
\label{eq:Hch}
\end{equation}
where, $\hat{\textbf{S}}_{k}$ is the spin-$1/2$ operator for dot $k$. $J_k$ denotes the exchange interaction between
$k$th and $(k+1)$th dots which is controlled using external gates.
The exchange couplings $J_k$ decrease exponentially with the distance between quantum dots \cite{burk99}, so, just nearest
neighbor interaction has been considered in Hamiltonian (\ref{eq:Hch}). Here $J_{k}>0$ ($\forall k\in [1,N-1]$) is for anti-ferromagnetic chain, while  $J_{k}<0$ ($\forall k\in [1,N-1]$) denotes the ferromagnetic chains.

Just as the initial proposal for the state transferring \cite{bose03} we consider an arbitrary state in the sender qubit (here spin $0$)
\begin{equation}
 |\psi_{in}\ra=\cos(\frac{\theta}{2})|0\ra+e^{i\phi}\sin(\frac{\theta}{2})|1\ra,
\end{equation}
where, $0\leq\theta\leq \pi$ and $0\leq\phi< 2\pi$ determine the location of the quantum state on the surface of the Bloch sphere.
The other spins are initialized in $|\psi_{ch}\ra$, the ground state of the Hamiltonian (\ref{eq:Hch}). The initial state of the system is thus
\begin{equation}
|\psi(0)\ra=|\psi_{in}\ra\otimes|\psi_{ch}\ra.
\label{eq:init}
\end{equation}
In the FM and also AFM chains with odd $N$ the ground state of the system is degenerate and
to choose a single state we add a small global magnetic field in the $z$ direction to break the symmetry and choose one of the ground states.
To send the state $|\psi_{in}\ra$ through the chain one can switch on the interaction between the $0$th and the $1$st spin of the
channel at $t=0$ as shown in Fig. \ref{fig0}(a). The interaction Hamiltonian takes the form
\begin{equation}
H_I=J_{0}\hat{\textbf{S}}_{0}\cdot\hat{\textbf{S}}_{1}.
\label{eq:Hint}
\end{equation}
So, the overall Hamiltonian is

\begin{equation}
H=H_{ch}+H_I.
\label{eq:Hperfect}
\end{equation}
Since, the initial state (\ref{eq:init}) is not an eigenvector of the Hamiltonian $H$, the whole system evolves as
\begin{equation}
 |\psi(t)\ra=\mathcal{T} e^{-i\int_{0}^{t}H(\tau)d\tau} |\psi(0)\ra,
\end{equation}
where, $\mathcal{T}$ denotes the time ordering operator and $\hbar$ has set to be one.
The time dependence of $H(\tau)$ may stem from random time dependent fluctuations. State of the receiver qubit $\rho_N(t)$ can be computed by tracing out
the other spins. So that we can define the channel $\xi$ as
$\rho_N(t)=\xi(\rho_0(0))$,
where, $\rho_0(0)=|\psi_{in}\ra \la\psi_{in}|$ is the density matrix of the input state at $t=0$.
To quantify the quality of the state transferring we compute the fidelity between
the sent and the received state $F(\theta,\phi;t)=\la \psi_{in} | \rho_N(t)|\psi_{in} \rangle$.
For a general quantum channel, including spin chains, we get
\begin{eqnarray}
 F(\theta,\phi;t)&=& \cos^4(\frac{\theta}{2})\langle0|\xi(|0\rangle\langle0|)|0\rangle
 + \sin^4(\frac{\theta}{2}) \langle1|\xi(|1\rangle\langle1|)|1\rangle\nonumber\\
 &+&\frac{1}{4}\sin^2(\theta) (\langle1|\xi(|0\rangle\langle0|)|1\rangle+\langle0|\xi(|1\rangle\langle1|)|0\rangle)\nonumber\\
 &+&\frac{1}{4}\sin^2(\theta) (\langle1|\xi(|1\rangle\langle0|)|0\rangle+\langle0|\xi(|0\rangle\langle1|)|1\rangle).
 \label{eq:Fthetaphi}
\end{eqnarray}

As it is clear from Eq. (\ref{eq:Fthetaphi}) this quantity is dependent on the initial state and we average over all possible input states, i.e. over the surface of the Bloch sphere,
to get an input independent quantity
\begin{equation}
 F_{av}(t)=\frac{1}{4\pi}\int_{\phi=0}^{\phi=2\pi}\int_{\theta=0}^{\theta=\pi} F(\theta,\phi;t) sin(\theta) d\theta d\phi.
\end{equation}

For an arbitrary channel we can write the average fidelity, $F_{av}$, in a simple general way
\begin{eqnarray}
F_{av} &=& \frac{1}{3}(\langle0|\xi(|0\rangle\langle0|)|0\rangle+\langle1|\xi(|1\rangle\langle1|)|1\rangle)\nonumber\\
&+&\frac{1}{6}(\langle1|\xi(|0\rangle\langle0|)|1\rangle+\langle0|\xi(|1\rangle\langle1|)|0\rangle)\nonumber\\
&+&\frac{1}{6}(\langle1|\xi(|1\rangle\langle0|)|0\rangle+\langle0|\xi(|0\rangle\langle1|)|1\rangle).
\label{eq:Fg}
\end{eqnarray}
Notice that in our case $F_{av}$ is a function of time $t$ and it takes its maximal value at a certain time $t=t_{opt}$.
The general form of fidelity (\ref{eq:Fg}) can be sensibly simplified by choosing a particular state for $|\psi_{ch}\ra$. For instance, in
the FM regime ($J_{k}<0$), the initial state of the channel is $|\psi_{ch}(0)\rangle = \prod_{k=1}^{N}|0\rangle_{k}$
and since the operator $S_z=\sum_{k=0}^{N}S_{k}^{z}$ commutes with the total Hamiltonian $H$ the
number of excitation is conserved at all times. Thus, evolution can be fully explained in the
subspace including the ground state $|\mathbf{0}\rangle = \prod_{k=0}^{N}|0\rangle$ and all single excitation
states $|1_j\rangle = \hat{\sigma}_{j}^{\dagger}|\mathbf{0}\rangle$ ($j=0,1,...,N$). The average fidelity of the ferromagnetic chain has been computed in Ref. \cite{bose03} as
\begin{equation}
 F_{av}^{FM} = \frac{1}{2} + \frac{|f_{N0}|^2}{6} + \frac{|f_{N0}|\cos(\gamma-\gamma_{0})}{3},
 \label{eq:Ffm}
\end{equation}
where, $f_{N0}(t)=\langle1_N|U(t)|1_0\rangle$ is the transition amplitude from spin 0 to the last one and
the phase of the transmission amplitude is fixed and defined as $\gamma(t) = \arg(f_{N0})$.
With a local unitary rotation to the $N$th spin, or equivalently applying a global magnetic field with a particular strength,
one can correct this phase. So, we have subtracted the phase $\gamma_0$ in the Eq. (\ref{eq:Ffm}) and ideally if
all parameters of the Hamiltonian are known we can tune $\gamma_0$ such that $\cos(\gamma-\gamma_{0})=1$ at optimal time $t=t_{opt}$.
But as our target here is to consider the effect of noise, the above condition cannot be met for arbitrary unknown disorder.

For AFM exchange interaction ($J_{k}>0$), when $N$ is even (channel has a unique ground state)
the effect of the channel is a fully symmetric depolarizing channel and all states are transmitted with equal fidelity \cite{bayat-xxz}.
So, transmission of any arbitrary state and its final fidelity specifies the average fidelity of even AFM chain
\begin{equation}
 F_{av}^{AFM} = \la 0| \xi(|0\ra \la 0|) |0\ra,
 \label{eq:Fafm}
\end{equation}
where, we have considered the transmission of state $|0\ra$.
Unfortunately, such compact results does not exist for AFM chains with odd $N$, however, Hamiltonian still have the symmetry of conserving the number of excitations.

Besides the quantum state transferring, one can consider entanglement distribution as well.
In this scheme, instead of sending a pure state through the spin chain we prepare
a singlet state between spin $0$ and and extra spin $0'$, shown in Fig. \ref{fig0} (b).
The rest of the system is again initialized in $|\psi_{ch}\ra$, the ground state of $H_{ch}$. At $t=0$,
spin $0$ is coupled to the chain (as it was in the state transferring strategy) while spin $0'$
remains decoupled during the evolution. As the result when the state of the $0$th spin goes
through the chain and reaches the last site we end up with an entangled state between
spin $0'$ and spin $N$. For a general channel $\xi$ the output state is:

\begin{eqnarray}
\rho_{0',N}(t)&=&\frac{1}{2}\{|0\rangle\langle 0|\otimes \xi(|1\rangle\langle 1|)+|1\rangle\langle 1|\otimes \xi(|0\rangle\langle 0|)\nonumber\\
&-&|1\rangle\langle 0|\otimes \xi(|0\rangle\langle 1|)-|0\rangle\langle 1|\otimes \xi(|1\rangle\langle 0|)\},
\label{eq:rho0N}
\end{eqnarray}
where, in each element the first part is the state of spin $0'$ and the second part represents the state of spin $N$.
The entanglement between $0^{'}$ and $N$ is usually measured by the concurrence
$C$ \cite{woot98}. For the FM case, the concurrence has a very simple form
\begin{equation}
C^{FM} = |f_{N0}(t)|.
\end{equation}

In the case of AFM chains with even $N$ again we have a simple form for the concurrence as
\begin{equation}
C^{AFM}=3\la 0| \xi(|0\ra \la 0|) |0\ra-2.
\end{equation}
In compare to Eq. (\ref{eq:Fafm}) we find a simple relationship between the average fidelity and the concurrence $C^{AFM}=3F_{av}^{AFM}-2$.

\section{Disordered Chains} \label{disorder}
In the previous section, we have considered an ideal situation in which there
is no disorder in the quantum dot chain. Experimental
\cite{koppen06,john05} and theoretical
results \cite{taylor07,khaet02,chris04,hu06,li10} show that the
hyperfine interaction and the exchange interaction fluctuations are the most significant
deleterious effects on quantum dot chains. Thus, it is very important to give a comparison of
state transferring performance between FM and AFM spin chains in the presence of these two practically important disorders.

\subsection{Hyperfine Interaction} \label{hyperfine}

For electron spins in quantum dots, the most important destructive phenomenon is interaction with the spin of nuclei in the bulk, i.e. hyperfine interaction.
In this part we study this effect on the quality of state transferring in both FM and AFM chains.

In the mesoscopic quantum dot systems, the electron spin interacts with many nuclear spins of its host material,
and it can be described by the Hamiltonian of the Fermi contact hyperfine
interaction \cite{hanson07,taylor07,paget77} as
$\mathcal{H}_{HF}=\sum\limits^{M}_{j=1}a_{j}\hat{\mathbf{I}}_{j}.\hat{\mathbf{S}}$, in which
$\hat{\mathbf{I}}_{j}$ denotes the spin of the $j$th nucleus, $\hat{\mathbf{S}}$ is the electron spin
operator and $a_{j}$ represents the coupling strength between the $j$th nucleus and the electron spin.
An alternative way to describe the average effect of nuclear spins is to treat them as an effective magnetic
field $\hat{\mathbf{B}}$, which is also called as the Overhauser field:
$(\sum\limits^{M}_{j=1}a_{j}\hat{\mathbf{I}}_{j})\cdot\hat{\mathbf{S}} = \hat{\mathbf{B}}\cdot\hat{\mathbf{S}}$.
Introducing the hyperfine interaction into spin chain system, the channel Hamiltonian $H_{ch}$ and total Hamiltonian $H$ are changed accordingly to
\begin{eqnarray}
H_{ch}^B&=&H_{ch}+\sum\limits_{k=1}^{N}\hat{\mathbf{B}}_{k}\cdot\hat{\mathbf{S}}_{k}, \cr
H^B&=&H+\sum\limits_{k=0}^{N}\hat{\mathbf{B}}_{k}\cdot\hat{\mathbf{S}}_{k},
\label{eq:HB}
\end{eqnarray}
where the nuclear field $\hat{\mathbf{B}}_{k}$ is a three-dimensional random vector and $J_k=J$ is assumed to be
constant for all quantum dots. Under the quasi-static approximation \cite{taylor07} the spin of nuclei do not change
in the state transferring time scale and $\hat{\mathbf{B}}_{k}$ is supposed to be
time independent. In the large $M$ limit, the random vectors $\hat{\mathbf{B}}_{k}$ have a Gaussian distribution \cite{taylor07}

\begin{equation}
P(\hat{\mathbf{B}})= \frac{1}{(2\pi B_{nuc}^{2})^{3/2}}\exp(-\frac{\hat{\mathbf{B}}\cdot\hat{\mathbf{B}}}{2 B_{nuc}^{2}}),
\label{eq:hdistri}
\end{equation}
with expectation value $\langle\hat{\mathbf{B}}\rangle = 0$ and standard deviation $B_{nuc}$.

Since, the hyperfine interaction term does not commute with $S_z$ it breaks the conservation of
spin-excitations so that we have to consider the total Hilbert space for the evolution which restricts
our simulation to rather short chains. We fix a random vector $\hat{\mathbf{B}}_{k}$ for each quantum dot
according to the distribution (\ref{eq:hdistri}) at $t=0$. Spin 0 is initialized to $|\psi_{in}\ra$ and the
channel is set to be in $\Pi_{k=1}^N|0\ra$ for FM chains and $|\psi_{ch}^B\ra$, the real ground state of $H_{ch}^B$, for AFM chains.
So, the initial state of the system is
\begin{equation}
|\psi(0)\ra=|\psi_{in}\ra\otimes|\psi_{ch}^B\ra.
\label{eq:initB}
\end{equation}
Then, we switch on the interaction between spin $0$ and spin $1$ and accordingly system evolves under action of the Hamiltonian $H^B$.
So, the average fidelity is computed for a fixed set of $\{\hat{\mathbf{B}}_{k} \}$ and since these are some random vectors
we have to average over hundreds of different realizations (we choose 500 times
in our simulations) of random vectors $\{\hat{\mathbf{B}}_{k} \}$ to get $\langle F_{av}\ra_B$ and $\langle C\rangle_B$.

\begin{figure} \centering
    \includegraphics[width=8.5cm,height=7.5cm,angle=0]{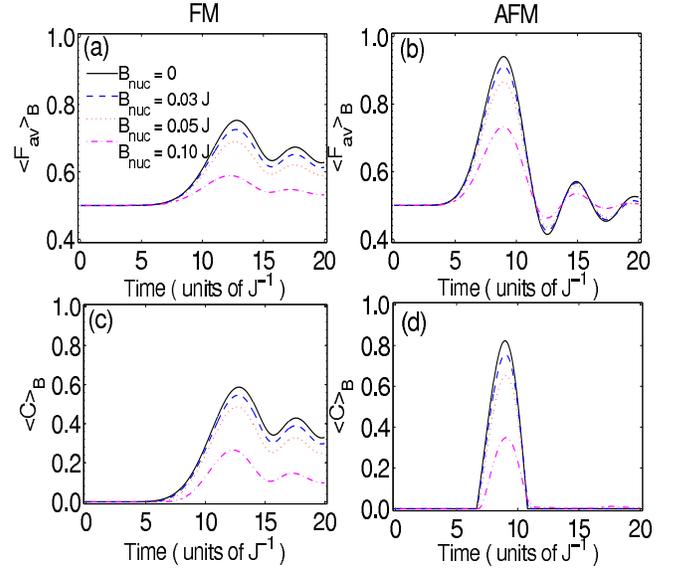}
    \caption{(Color online) Time evolutions of the average fidelity $\langle F_{av}\rangle_{B}$ as well as the
concurrence $\langle C\rangle_{B}$ in a chain of length $N=10$ in the presence of hyperfine interaction for FM (Figs. (a) and (c)) and
AFM quantum dot chains (Figs. (b) and (d)).
Here $J$ is absolute value of the exchange coupling between two dots. }
     \label{fig1}
\end{figure}

In Fig. \ref{fig1} we show the evolutions of average fidelity $\langle F_{av}\rangle_{B}$ and
concurrence $\langle C\rangle_{B}$, which exhibit the performance of quantum information transferring,
for FM and AFM quantum dot chains of length $N = 10$. As it is clear from
Fig. \ref{fig1}, the effect of the hyperfine interaction is always destructive and decreases
the quality of classical transmission such that the stronger the hyperfine interaction, the lower the quality of transmission. The average optimal time, where the peak of $\langle F_{av}\rangle_{B}$ and $\langle C\rangle_{B}$ locate,
is the same and does not change with increasing the strength of the hyperfine interaction.
Notice that the optimal time for each realization of the chain might be different due to the random nature of disorder, but since we do not know how disorder changes the Hamiltonian, we cannot modify it according to the disorder, and we only can consider its average value which our simulations show that it is not affected by disorder after many trials. Another feature of the Fig. \ref{fig1} is the fact that in the presence of disorder the first peak becomes the dominant peak in the evolution and however, the subsequent peaks may be higher for an ideal
situation without disorder but in the presence of disorder one can concentrate just on the first peak, as we will do in the rest of the paper.
The most significant results of the Fig. \ref{fig1} come from the
comparison between FM and AFM chains. According to all quantities shown in Fig. \ref{fig1} the quality of
transmission in AFM chains is always higher and they give a higher value in their peak and a lower optimal
time which peak occurs. Having a faster dynamics in the AFM regime is very important because disorder and all
other decoherence sources have less opportunity to interfere with the evolution.

\begin{figure} \centering
    \includegraphics[width=8.5cm,height=7.5cm,angle=0]{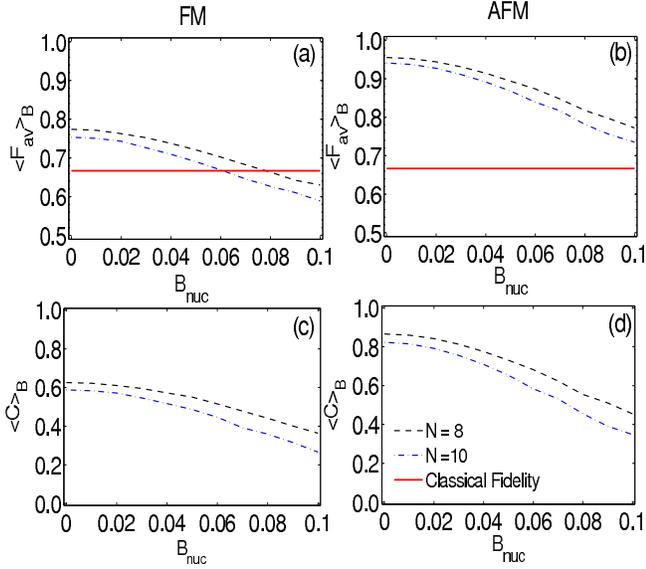}
    \caption{(Color online) Comparison of average fidelity $\langle F_{av}\rangle_{B}$ and concurrence
$\langle C\rangle_{B}$ between FM (Figs. (a) and (c)) and AFM (Figs. (b) and (d)) quantum dot chains
in terms of standard deviation $B_{nuc}$ for different lengths.
The red straight solid line represents the highest average fidelity accessible to the classical teleportation scheme.}
     \label{fig2}
\end{figure}

In Fig. \ref{fig2} we have plotted $\langle F_{av}\rangle_{B}$ and $\langle C\rangle_{B}$ in terms
of hyperfine interaction strengths $B_{nuc}$ in FM and AFM chains of length $N = 8$ and $N = 10$.
As discussed above we just consider $t=t_{opt}$ where $t_{opt}$ is the time in which the first peak occurs.
As it has been shown in Fig. \ref{fig2} (a)
for FM chains the average fidelity decreases very slowly for small values of $B_{nuc}$
and becomes less than the average fidelity of classical teleportation, which is
equal to $2/3$, when $B_{nuc}=0.08J$ ($B_{nuc}=0.06J$) for the chain of length $N=8$ ($N=10$).
For AFM chains even for $B_{nuc}=0.1J$ (which is a very pessimistic estimation) average fidelity is still above
the classical threshold limit for the same length.
Again Fig. \ref{fig2} shows that the quality of communication in AFM chains are better than FM ones according to
 both $\langle F_{av}\rangle_{B}$ and $\langle C\rangle_{B}$. Particularly, small amount of disorder
($B_{nuc} < 0.01 J$) almost does not change the quality of communication.

\begin{figure} \centering
    \includegraphics[width=8.5cm,height=7.5cm,angle=0]{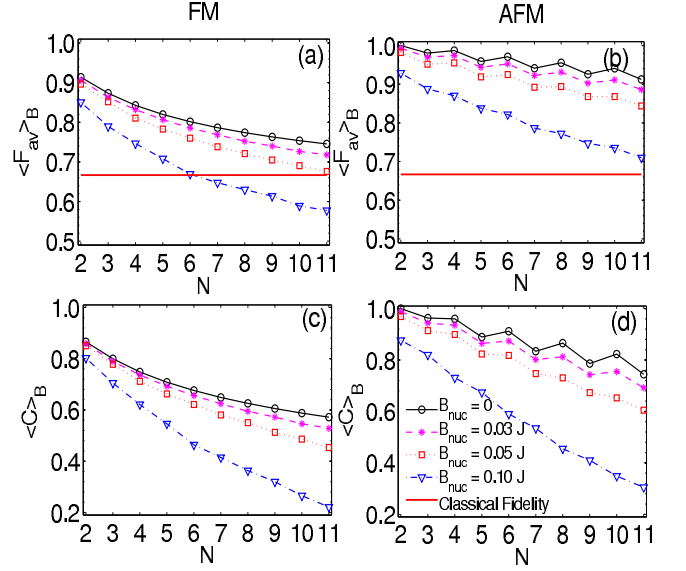}
    \caption{(Color online) Comparison of average fidelity $\langle F_{av}\rangle_{B}$ and concurrence $\langle C\rangle_{B}$ in
FM (Figs. (a) and (c)) and AFM  (Figs. (b) and (d)) quantum dot chains as a function of length $N$ for different values of $B_{nuc}$.
The red straight solid line represents the highest average fidelity accessible to the classical teleportation scheme.}
     \label{fig3}
\end{figure}

In Fig. \ref{fig3} we show the performance of $\langle F_{av}\rangle_{B}$ and $\langle C\rangle_{B}$ in terms
of length for some fixed $B_{nuc}$ in both FM and AFM chains. For FM chains, the average fidelity decreases by
increasing the length and when $B_{nuc}=0.1J$ (which is quite a pessimistic estimation) for chains up to $N=6$ we can transfer our information better
than 2/3, highest average fidelity of classical communication.
For $B_{nuc} < 0.05$ we are above the classical threshold $2/3$ even for a
chain of $N = 11$ spins. For AFM chains, as it has been shown in Fig. \ref{fig3}(b) even when $B_{nuc}$ is
very strong we are above the classical threshold for a chain of $N = 11$ and for more reasonable values of $B_{nuc}$ we are
far beyond the fidelity of $2/3$. Here, due to the different symmetry in the ground state of the even and odd chains
we have an even-odd effect and even chains give a higher quality in their transmission.
This even-odd effect can be seen through the zigzag behavior of the average fidelity and entanglement in AFM chains.
We also have considered the concurrence as a function of length $N$ for chains of different $B_{nuc}$ in Figs. \ref{fig3} (c) and (d).
As it was expected we found a higher entanglement in AFM chains with the same length than the FM ones and
similar to the average fidelity we have even-odd effect for concurrence in AFM chains. This is an extension of the results
for non-disordered chains presented in \cite{bayat-AFM, bayat-xxz}.

\subsection{Exchange Coupling Fluctuations}
\label{fluctuation}
In order to successfully accomplish state transferring before the relaxation time of electron spins,
the information propagation speed should be fast and exchange interaction need to be strong.
Exchange interaction in a chain of quantum dots can be easily controlled by gate voltages.
However, using external gates to control exchange interactions would inevitably introduce background
charge fluctuation in the environment. The deleterious effect of charge fluctuations on the quantum dot
chains mainly has two aspects: (i) generating variations in the barrier heights; (ii) causing a
random bias potential between the neighboring dots.
Consequently, exchange couplings $J_k$ in gated quantum dots unavoidably fluctuates with background charge
fluctuation such that spin qubits in quantum dot chain suffer dephasing \cite{burk99,hu06,li10}.

To simulate the effect of these fluctuations on the quality of transmission we consider
the couplings between neighboring dots as $J_k=J(1+\delta_k(t))$.
The dimensionless parameters $\delta_k(t)$ are time-dependent random variables
and have two main properties: (i) disorder in each site is independent from the other
sites; (ii) in each site $k$, the disorder parameter $\delta_k(t)$ is correlated in time such
that the frequency spectrum behaves as $S(f) = \sigma_{J}/f^{\alpha}$, where $\sigma_{J}$ denotes
the standard deviation and $\alpha$ defines the type of the noise. For instance, $\alpha=0$ represents
the white noise, $\alpha=1$ denotes the $1/f$ noise (pink noise),  $\alpha=2$ is known as the
Brownian noise and finally $\alpha=\infty$ is the static noise. In appendix A we have
given a method to generate $\delta_k(t)$ such that their frequency spectrum
behaves as $S(f) = \sigma_{J}/f^{\alpha}$. In Ref. \cite{chris04} it was shown that the
fluctuations of the coupling in a quantum dot chain mainly behaves like $1/f$ noise (pink noise).
In our simulation we consider the following Hamiltonians for initializing the system
\begin{eqnarray}
H_{ch}^J&=&H_{ch}, \cr
H^J(t)&=& \sum\limits_{k=0}^{N-1}J(1+\delta_k(t))\hat{\textbf{S}}_{k}\cdot\hat{\textbf{S}}_{k+1}.
\label{eq:HJ}
\end{eqnarray}
We do not consider any noise effect in the channel Hamiltonian, which simply means that for
both FM and AFM chains we always consider an ideal state for the channel.
It means that we take the state given in Eq. (\ref{eq:init}) as the initial state of the system.
The reason that we ignore fluctuations in the initial state comes from the fact that in
FM chains these fluctuations do not change the ground state and in the AFM
chains when we consider an static random fluctuations even up to $\sigma_J=0.1J$ the fidelity between
the real ground state and the ideal (without disorder) ground state is always above $0.99$ for all lengths that we have considered in this paper.

For evolving the system, at each time step we generate $\delta_k(t)$ according to their frequency
spectrum $S(f)$ and system evolves according to Hamiltonian $H^J(t)$. The important issue about
this particular Hamiltonian is that it does not break the symmetry of the system and Hamiltonian still commutes with $S_z$.
Consequently, the average fidelity preserves its form of Eq. (\ref{eq:Ffm}) for
FM chains and Eq. (\ref{eq:Fafm}) for even AFM chains, just like the other quantities,
i.e. entanglement and excitation transmission amplitudes. The only difference is the
fact that the parameters in those formula are not deterministic anymore and they are random.
So, similar to hyperfine interaction we average over many realization of coupling disorders
to get average fidelity $\la F_{av}\ra_J$ and entanglement $\la C\ra_J$.

\begin{figure} \centering
    \includegraphics[width=8.5cm,height=7.5cm,angle=0]{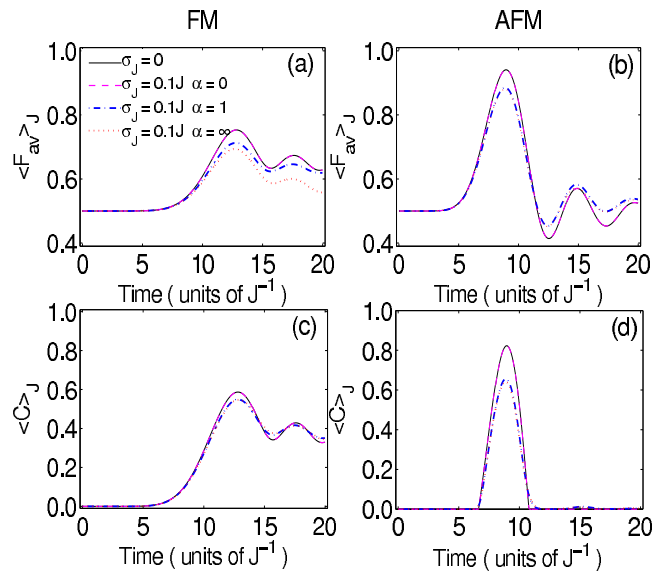}
    \caption{(Color online) Comparison of average fidelity $\langle F_{av}\rangle_{J}$ and concurrence
$\langle C \rangle_{J}$ between FM (Figs. (a) and (c)) and AFM (Figs. (b) and (d)) chains of length $N = 10$ as
a function of time for $\sigma_{J} = 0.1 J$
and different sort of noises such as $\alpha=0$ (white noise), $\alpha = 1$ ($1/f$ noise) and $\alpha = \infty$ (static noise).
Here $J$ is absolute value of the exchange coupling between two dots.}
     \label{fig4}
\end{figure}

In Fig. \ref{fig4}, we have plotted the average fidelity $\langle F_{av}\rangle_{J}$ and concurrence
$\langle C \rangle_{J}$ for FM and AFM chains of length $N = 10$ in terms of time $t$ in the presence of exchange coupling fluctuations. Here we consider three kinds of exchange coupling noises with $\langle {\delta_k}\rangle=0$ and standard
deviation $\sigma_{J} = 0.1 J$: white noise with Gaussian distribution, $1/f$ noise
and the static noise, again with Gaussian distribution. We find that $\langle F_{av}\rangle_{J}$ and
$\langle C \rangle_{J}$ do not change under the action of white noise.
This can be explained in the way that $\delta_k(t)$ is a random variable which is independent at different times so, its effect
is compensated at different time steps such that in average it does not affect the communication scheme at all. As it is clear from Fig. \ref{fig4}, the effect of $1/f$ noise on FM chains is quite similar to the static noise. These results show that the faster the
exchange coupling changes, the higher the fidelity and concurrence of state transfer through the quantum dot chain.
It is worthwhile to say that the optimal time for  $\langle F_{av}\rangle_{J}$ and
$\langle C \rangle_{J}$, also does not change for exchange coupling fluctuations. However, for each realization the optimal time might be different but since those changes are random we do not have any prior knowledge about them and we have to take the average optimal time which is fully independent of disorder in the limit of large number of trials.

\begin{figure}
\centerline{\includegraphics[width=8.5cm,height=7.5cm,angle=0]{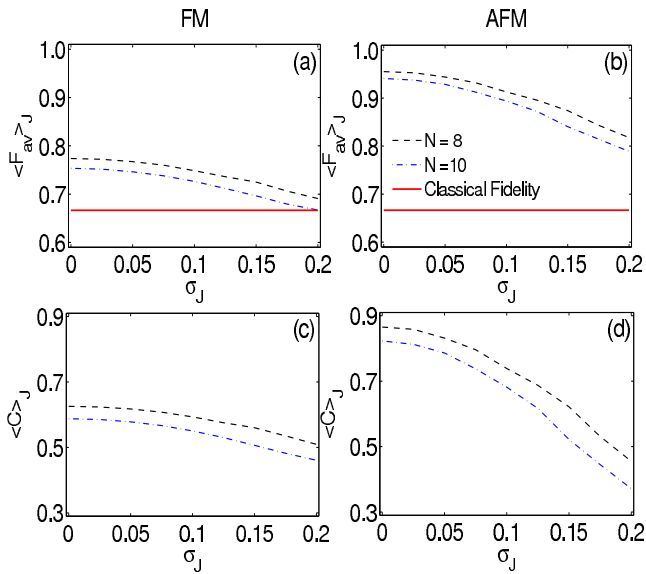}}
\caption{(Color online) Average fidelity $\langle F_{av}\rangle_{J}$ and concurrence
$\langle C \rangle_{J}$ in terms of disorder strength $\sigma_J$ in both FM (Figs. (a) and (c)) and AFM chains (Figs. (b) and (d)) for different lengths.
The red straight line at $2/3$ shows the fidelity accessible to the classical teleportation scheme.}
\label{fig5}
\end{figure}

In Fig. \ref{fig5}, we have shown the average fidelity $\langle F_{av}\rangle_{J}$ and concurrence
$\langle C\rangle_{J}$  versus the standard deviation $\sigma_{J}$ in the presence of $1/f$ noise.
$\langle F_{av}\rangle_{J}$ and $\langle C\rangle_{J}$ decrease as the strength $\sigma_{J}$ increases.
For both FM and AFM chains, the average fidelity $\langle F_{av}\rangle_{J}$ is always beyond the classical threshold
$2/3$ for the length $N = 10$ even a disorder as strong as $\sigma_{J} = 0.2J$. In comparison to FM chains, AFM chains
have higher average fidelity and concurrence. For instance,
in the case of AFM chains, $\langle F_{av}\rangle_{J}=0.8$ while for FM chain it is $\langle F_{av}\rangle_{J}=0.66$
in a chain of length $N = 10$ and $\sigma_{J} = 0.2J$.

\begin{figure}
\centerline{\includegraphics[width=8.5cm,height=7.5cm,angle=0]{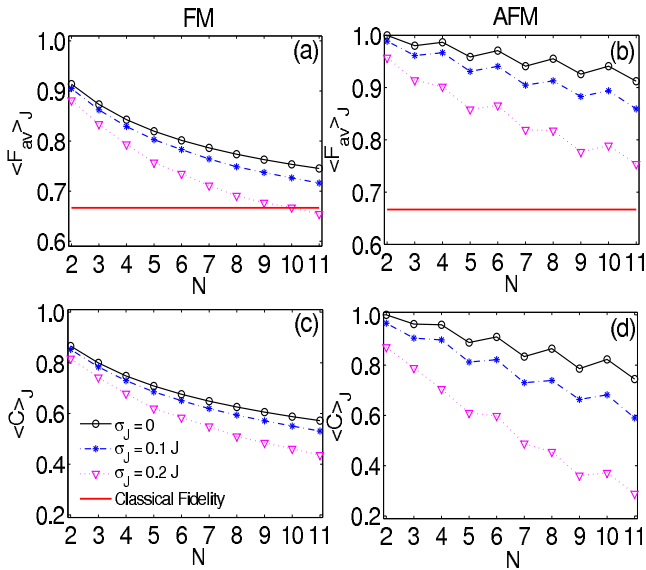}}
\caption{(Color online) Average fidelity $\langle F_{av}\rangle_{J}$ and
concurrence $\langle C \rangle_{J}$ versus the length $N$ in both FM (Figs. (a) and (c)) and AFM (Figs. (b) and (d))
chains for different values of $\sigma_J$.
The red straight line at $2/3$ shows the fidelity accessible to the classical teleportation scheme.}
\label{fig6}
\end{figure}

In Fig. \ref{fig6}, we give the simulation results for $\langle F_{av}^J\rangle$ and $\langle C\rangle_{J}$ in terms of
length $N$ when considering $1/f$ noise. The even-odd effect of AFM chain also
create the non-monotony evolutions of $\langle F_{av}\rangle_{J}$ and $\langle C\rangle_{J}$ with
respect to $N$. As Fig. \ref{fig6} shows,
for very pessimistic situation $\sigma_{J} = 0.2 J$, the average fidelity $\langle F_{av}\rangle_{J}$ in FM chains of $N = 10$ is
equal to the classical average fidelity 2/3, while $\langle F_{av}\rangle_{J}$ in AFM chains of $N = 11$ can achieve $0.75$.

\section{Realistic Scenario for Quantum State Transferring} \label{practical}

\begin{figure}
\centerline{\includegraphics[width=8.5cm,height=7.5cm,angle=0]{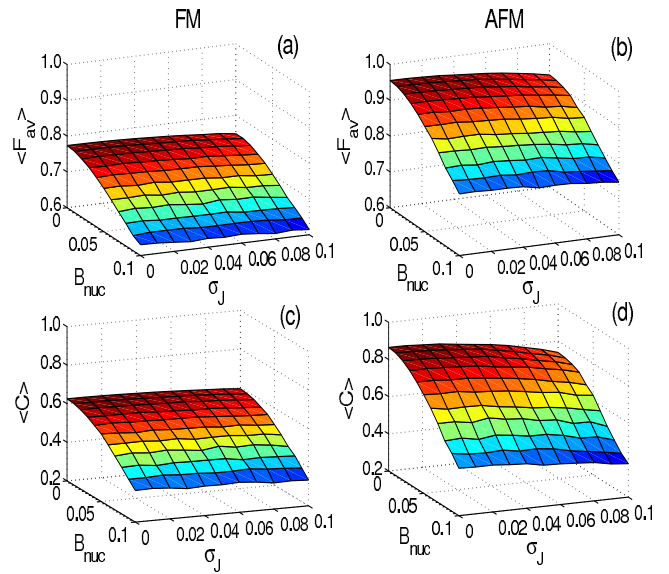}}
\caption{(Color online) Average fidelity $\langle F_{av}\rangle$ and concurrence
$\langle C \rangle$ in FM (Figs. (a) and (c)) and AFM (Figs. (b) and (d)) noisy
chains versus both $B_{nuc}$ and $\sigma_{J}$ in a chain of length $N = 8$.}
\label{fig7}
\end{figure}

In a practical case we suffer from both main sources of noise simultaneously, i.e. hyperfine interaction and exchange coupling fluctuations.
We have considered both of these noises together in Fig. \ref{fig7}.
As it can be seen from Fig. \ref{fig7}, both the hyperfine interaction and exchange coupling fluctuations
give a destructive impact on quantum information transmission.  Comparing the influence of hyperfine
interaction and exchange coupling fluctuation on average fidelity $\langle F_{av}\rangle$ and
concurrence $\langle C\rangle$, we find that the hyperfine interaction is more destructive to state transfer
than exchange coupling noise and AFM chains is more robust against disorders than FM ones.
The reason that the hyperfine interaction is more destructive is due to the fact that it breaks the symmetry of the system and
changes the number of excitations during the evolution. This put the system out of the subspace of the initial state and gives more destructive result.
Another important point to note is the fact that for the even AFM channels the output state $\rho_{0'N}$ remains a Werner state (a mixed state in which
the singlet is mixed with identity) even in the presence of the hyperfine noise as it is in a random direction.
For example, by averaging over 500 different noise profiles we found that the deviation from the Werner state at
optimal time is less than $0.1\%$ according
to the matrix elements.
As these states allow entanglement
distillation according to known protocols \cite{Bennett96} using even AFM quantum dot chains should be highly desirable.

\begin{figure}
\centerline{\includegraphics[width=8.5cm,height=3.5cm,angle=0]{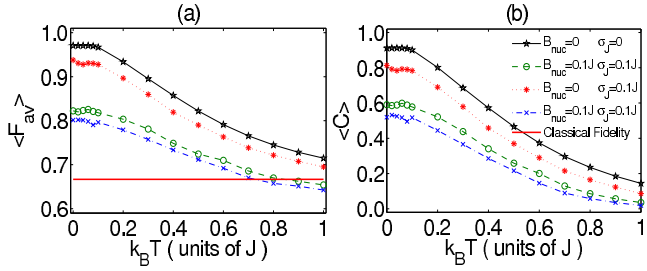}}
\caption{(Color online) Average fidelity $\langle F_{av}\rangle$ (Fig (a)) and concurrence
$\langle C \rangle$ (Fig (b)) in AFM
chains versus temperature $T$ in a chain of length $N = 6$ in the presence of disorder.}
\label{fig8}
\end{figure}

Another challenging problem for implementing quantum state transfer in the laboratory is initializing the system to its ground state.
It has been shown that in the limit of large $N$ cooling the system to its ground state takes an exponentially long time \cite{gottesman09,Kay09}.
This is truly an important problem for gapless systems, such as ours, which the energy separation between the ground state and the excited states vanishes for long chains and approaching the ground state adiabatically becomes challenged. However, in our scheme we consider only
finite chains and there is always an energy gap between the ground state and the excited states manifold. If we can prepare the system in a temperature $T$ such that its thermal energy $k_BT$, where $k_B$ is the Boltzmann constant, is less than its energy gap then, system is well explained by its ground state. Otherwise, in thermal equilibrium at temperature $T$ the initial state of the channel is described by
\begin{equation}
\rho_{ch} = \frac{\exp(-H_{ch}/(k_{B}T))}{\mathrm{Tr}[\exp(-H_{ch}/(k_{B}T))]}.
\end{equation}
We note that $H_{ch}$ should be replaced by $H_{ch}^{B}$ in the case of having hyperfine interaction. In Fig. \ref{fig8}, we have plotted the average fidelity $\langle F_{av}\rangle$ and concurrence $\langle C \rangle$ as functions of temperature $T$ in noiseless and different disordered AFM chains for length $N = 6$. It is shown that hyperfine interaction, exchange interaction fluctuation and increasing temperature are always the deleterious effects on quantum state transfer, and the hyperfine interaction is more destructive to system than exchange interaction fluctuation. Moreover, the evolutions of $\langle F_{av}\rangle$ and $\langle C \rangle$ versus thermal energy $k_{B}T$ show a plateau in the regime of $k_{B}T \leq 0.1 J$, before going down for $k_{B}T > 0.1 J$. This width of this plateau shows the energy gap between the ground state and first excited state for the finite spin chain. If the thermal energy $k_{B}T$ is much smaller than the energy gap, it is unlikely to populate excited states so that the system remains in its ground state. For AFM chain of length $N = 6$, when both hyperfine interaction $B_{nuc} = 0.1 J$ and exchange coupling $\sigma_{J} = 0.1J$ are taken into account, the average fidelity $\langle F_{av}\rangle$ is beyond the classical fidelity $2/3$ for $k_{B}T<0.7 J$, and $\langle F_{av}\rangle$ and $\langle C \rangle$ is beyond $0.79$ and $0.49$ respectively at $k_{B}T=0.1 J$. Here we only consider the thermal effect in AFM chains, since it has been reported that AFM chains performs better than FM ones in quantum state transfer under the thermal fluctuations \cite{bayat-xxz}.

\section{Exploiting Optimal Control Theory For Improving the Results} \label{optimal}

\begin{figure} \centering
    \includegraphics[width=8.5cm,height=7.5cm,angle=0]{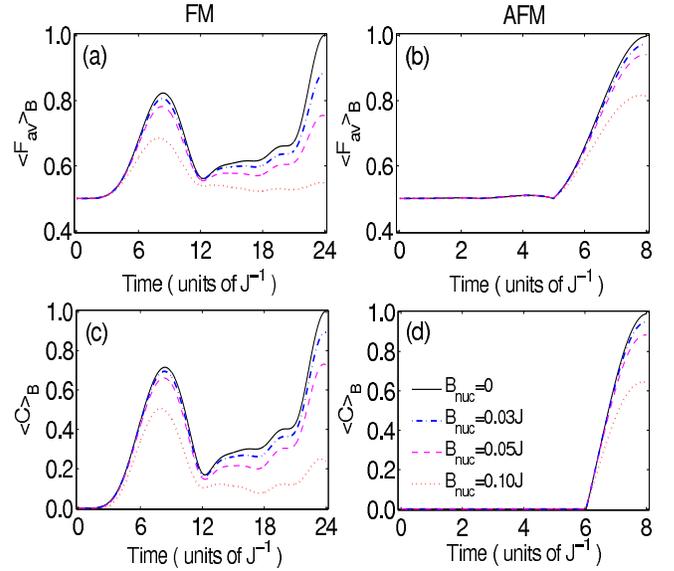}
    \caption{(Color online) Time evolution of average fidelity $\langle F_{av}\rangle_{B}$ and concurrence $\langle C\rangle_{B}$ in
FM (Figs. (a) and (c)) and AFM  (Figs. (b) and (d)) quantum dot chains with optimal control for different values of $B_{nuc}$.}
     \label{fig9}
\end{figure}

\begin{figure} \centering
    \includegraphics[width=8.5cm,height=7.5cm,angle=0]{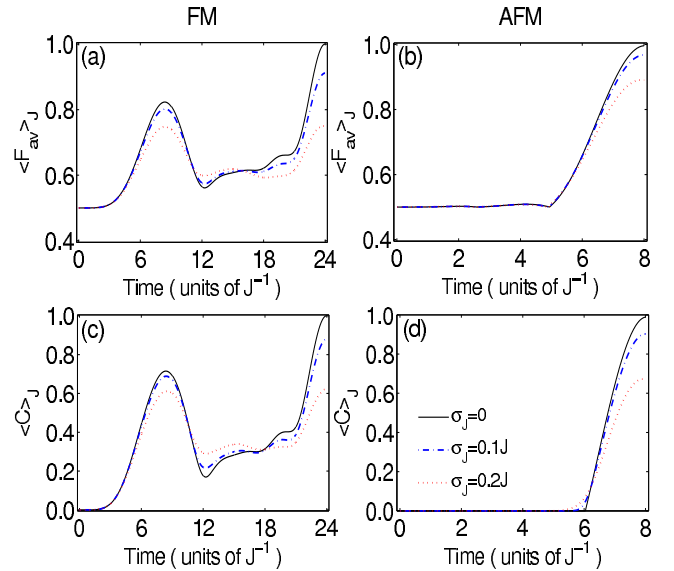}
    \caption{(Color online) Time evolution of average fidelity $\langle F_{av}\rangle_{J}$ and concurrence $\langle C\rangle_{J}$ in
FM (Figs. (a) and (c)) and AFM  (Figs. (b) and (d)) quantum dot chains with optimal control for different values of $\sigma_{J}$.}
     \label{fig10}
\end{figure}

An important question at this stage is whether the couplings in the chain can be tuned to certain values for maximizing the fidelity of state transfer.
To a certain extent, it should be possible to tune the barriers between the dots using electrostatic gates and thereby tune the Heisenberg interactions between spins.
It is however known that the engineering of static couplings cannot be used in a Heisenberg chain for taking the fidelity to unity even in the absence of disorder
\cite{Marcin08}. Thus, one may modulate one coupling in time and think of using optimal control theory to achieve the maximal fidelity. Optimal control theory uses a time-dependent pulse optimized to drive a system from a certain initial state to the target state. In this section we use the optimal control theory to improve the output while the initialization process is the same as before.
We simply modulate the coupling $J_{0}(t)$ between spin 0 and spin 1, just as used in \cite{schirmer08,bayat-control}, such that the perfect quantum state transfer can be achieved in Quantum dot chain at a target time $t_{f}$. We numerically search for the minimal value of $t_f$ to have a fast dynamics and give less opportunity to disorder and external noise for their destructive effect.
In an ideal situation, in the absence of disorder, we choose $J_{0}$ as the piecewise constant controls that can be simply approximated by square pulses which are preferable in practical situation. We divide the time interval $[0, t_{f}]$ into $k$ equal parts, and let $J_{0}$ be a constant value in each subinterval. Given a sequence of control pulses, the Hamiltonian of the system becomes as $H = H_{I} + H_{ch}$, where the control Hamiltonian is
\begin{equation}
H_{I} = J_{0}(t)\hat{\textbf{S}}_{0}\cdot\hat{\textbf{S}}_{1},
\end{equation}
and $J_{0}(t)$ takes a constant value in each time subinterval. Here, we adopt optimization based on quasi-Newtonian method to numerically generate the sequence of control pulses for FM and AFM chains of length $N=6$. We set the time steps to $k = 50$, and attempt to find the optimal set of $J_0(t)$ to maximize the average fidelity and concurrence at a minimum target time $t_{f}$. We have plotted the results in Fig. \ref{fig9}. We find that in noiseless FM (AFM) chains of length $N=6$, it is indeed possible to implement state transfer with almost unit (above 0.99) average fidelity and concurrence.
The minimal required time is found to be very different in FM ($t_{f} = 24$) and AFM ($t_{f} = 8$) chains.

Comparing the required time $t_f$ for FM and AFM chains shows a big advantage for AFM chains due to their fast dynamics. This advantage will be clear when we consider disorder in our setup. Fig. \ref{fig9} shows the time evolution of average fidelity $\langle F_{av}\rangle$ and concurrence $\langle C \rangle$ under different hyperfine interactions for FM and AFM chains when $J_0(t)$ varies according to its optimized pulse for the ideal situation. It is shown that AFM chain is more robust in the presence of hyperfine interaction. Comparing the results of Fig. \ref{fig3} and Fig. \ref{fig9} for a FM chain when  $B_{nuc} = 0.1$ shows that without optimizing any coupling we have $\langle F_{av}\rangle_{B} = 0.67$ and $\langle C \rangle_{B} = 0.46$ while, using optimal pulse gives $\langle F_{av}\rangle_{B} = 0.55$ and $\langle C \rangle_{B} = 0.25$. This means that in the presence of a strong hyperfine interaction the optimal control is not effective for FM chains and even gives lower values of $\langle F_{av}\rangle_{B}$ and $\langle C \rangle_{B}$ in compare to the simple methodology used in previous sections. This is because dynamics is so slow ($t_f$ is large) that disorder has enough opportunity to deteriorate the output quality and effectively there is no gain in using optimization. In contrast, in AFM chains even when $B_{nuc}$ is very strong optimization improves the results. This is because the target time $t_f$, needed for optimization process, is comparable with the time needed for ordinary transmission without optimization.

We can also consider the effect of exchange coupling fluctuations in the optimized coupling strategy.
In Fig. \ref{fig10} we have plotted the time evolution of $\langle F_{av}\rangle$ and $\langle C \rangle$ in the presence of exchange coupling fluctuations. As we expect, AFM chains behave better than FM chain against exchange coupling fluctuations, and the deteriorative effect of exchange coupling fluctuations is not as serious as hyperfine interaction. For this kind of disorder comparing our results for optimized coupling and non-optimized one shows that even for a strong disorder, $\sigma_J=0.2$, the optimization can improve the output quite significantly for both FM and AFM chains.

\section{Conclusion} \label{conclusion}

In summary, we have considered two inevitable types of disorders in quantum dot arrays for quantum communication, i.e. hyperfine interaction and
exchange coupling fluctuation.
We have considered quantum information transmission through the chain in both FM and AFM phases.
Our results show that disorder always has a destructive effect on
the quality of transmission however,
the AFM chains are much more resistive against disorder in the array of quantum dots than the FM ones.
In addition, AFM chains remain depolarizing channels in the presence of disorders which makes them
useful for entanglement distillation.
Rough verdict of the paper is that it is possible to use chains up to 10 quantum dots for quantum communication with fidelity exceeding 0.9 for AFM even
in the presence of realistic noises.
The average optimal communication time does not change with
disorder and also it was shown that
hyperfine interaction is more destructive than the exchange coupling fluctuations. This is due to the fact that
hyperfine interaction breaks the symmetry of conserving the number of excitations and consequently decoheres the quantum information more.
Furthermore, we have shown that quantum communication can be done robustly for thermal energies below the energy gap in a finite spin chain.

Finally, we showed that it is possible to improve the results with modulating the first coupling in time by the means of optimal control theory. However, because of a longer time needed for optimization, this strategy is not practically effective in FM chains when hyperfine interaction is strong.

{\em Acknowledgments.} We thank Xiaoting Wang for his valuable comments and in particular for his help in optimization part.
SB is supported by the EPSRC ARF grant EP/D073421/1,
through which AB is also supported. SB also acknowledges the Royal Society and the Wolfson Foundation.
SY is supported by CSC, National
Fundamental Research Program (Grant No. 2009CB929601), also by
National Natural Science Foundation of China (Grant No. 10674128 and
60121503) and the Innovation Funds and Hundreds of
Talents program of Chinese Academy of Sciences
and Doctor Foundation of Education Ministry of China (Grant No.
20060358043).

%%%%%%%%%%%%%%%%%%%%%%%%%%%%%%%%%%%
%%  APPENDICES %%%%%%%%%%%%%%%%%%%%%%%%%
%%%%%%%%%%%%%%%%%%%%%%%%%%%%%%%%%%%
%\newpage
\appendix

\section{$1/f^{\alpha}$ noise generation\label{s:noise}}

\begin{figure}
\centerline{\includegraphics[width=8cm,height=6cm,angle=0]{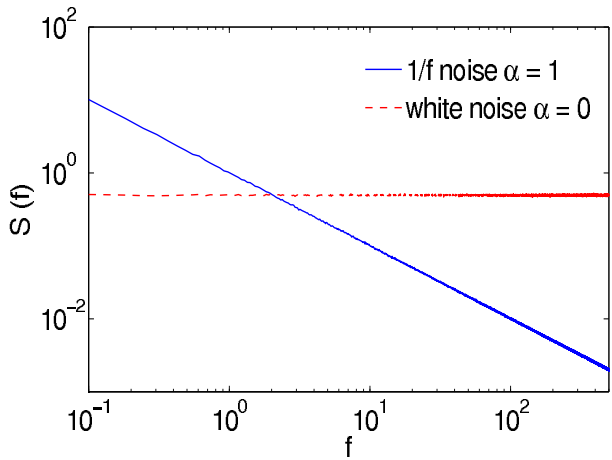}}
\caption{(Color online) The spectral density $S(f)$ over 1000 realizations for white
noise ($\alpha = 0$, red dashed line) and $1/f$ noise ($\alpha = 1$, blue solid line).
We have set $f_{max} = 1000$ and $M = 2^{14}$.}
\label{fig11}
\end{figure}

Here we apply Inverse Discrete Fourier Transform (IDFT) method to generate $1/f^{\alpha}$ noise \cite{jack00}.
The frequency spectrum is $S(f) = \sigma_J/f^{\alpha}$ where, $\sigma_J$ is the variance and $\alpha$ denotes the type of the noise.
For instance $\alpha = 0$ represents the white noise
while $\alpha = 1$ is for $1/f$ noise. The IDFT of $S(f)$ is defined as

\begin{equation}
s(t) = \frac{1}{M} \sum \limits ^{M-1}_{k=0} S(f_{k}) \mathrm{e}^{i 2\pi (f_{k}-\eta_{k})t},
\label{eq:Jnoise}
\end{equation}
where, $\eta_{k}$'s are independent random variables with mean 0 and variance 1 and $f_k = \frac{k}{M} f_{max}$ is
the discrete frequency between 0 and some numerical upper bound $f_{max}$. To show that Eq. (\ref{eq:Jnoise})
produces $1/f^\alpha$ noise we can generate $s(t)$ and then compute its
fourier transform according to deterministic frequencies. Since we have random variables $\eta_k$ in the Eq. (\ref{eq:Jnoise}) one can
repeat the process over hundreds of times (here we have done it
for 1000 times) and make the average. The results have been shown in Fig. \ref{fig11} for $\alpha=0$ (white noise)
and $\alpha=1$ ($1/f$ noise) in the logarithmic scale.  Fig. \ref{fig11} clearly
shows that the signal $s(t)$, given in Eq. (\ref{eq:Jnoise}), generates the desirable frequency spectrum $S(f)$.

\end{document}